\documentstyle[12pt,epsf]{article}
\renewcommand{\bar}[1]{\overline{#1}}

\def\ru1{\rule[-0.4truecm]{0mm}{1truecm}}

\begin{document}

\begin{flushright}
{\footnotesize 
BIHEP-TH-97-14\\ USM-TH-72}
\end{flushright}

\bigskip\bigskip
{\centerline{\Large
\bf New Sum Rules for Nucleon Tensor Charges\footnote{\baselineskip=13pt
Work partially supported 
by National Natural
Science Foundation of China
under Grant No.~19605006, by Fondecyt (Chile) under grant
1960536, and by a C\'atedra Presidencial (Chile).}}}

\vspace{22pt}

\centerline{
\bf 
Bo-Qiang Ma$^{a}$ and Ivan Schmidt$^{b}$}

\vspace{8pt}
{\centerline {$^{a}$CCAST (World Laboratory), P.O.~Box 8730, Beijing
100080, China}}

{\centerline {and Institute of High Energy Physics, Academia Sinica,
P.~O.~Box 918(4),}

{\centerline {Beijing 100039, China}}

{\centerline {e-mail: mabq@bepc3.ihep.ac.cn}}

\vspace{8pt}
{\centerline {$^{b}$Departamento de F\'\i sica, Universidad T\'ecnica Federico
Santa Mar\'\i a,}}

{\centerline {Casilla 110-V, 
Valpara\'\i so, Chile}}

{\centerline {e-mail: ischmidt@fis.utfsm.cl} }


\vspace{10pt}
\begin{center} {\large \bf Abstract}

\end{center}

Two new sum rules for the quark tensor charges
of the nucleon are proposed, based on a relation connecting
the quark transversity distributions to the
quark helicity distributions and the quark model spin distributions,
and on the sum rules for the quark helicity distributions.
The two sum rules are
useful for an estimate of the values of
the quark tensor charges
$\delta U$ and $\delta D$
from the measured quantities of $\Gamma^p$, $\Gamma^n$,
$g_A/g_V$ and
$\Delta S$, and two model correction factors with limited
uncertainties.
We predict a small value for the sum of the
quark tensor charges compared to most other predictions,
in analogy
to the
unexpectedly small quark helicity sum
which gave rise to the proton ``spin puzzle''.

\vfill
\centerline{
}
\vfill

\newpage

Historically,
parton sum rules have played important roles
in the understanding of the quark-gluon structure of
the nucleons. The confirmation
of
the Gross-Llewellyn~Smith (GLS) \cite{GLSSR},
Gottfried \cite{GSR}, and Adler \cite{ASR} sum rules
in early deep inelastic scattering (DIS)
experiments on unpolarized
structure functions were important for identifying
the quantum numbers of partons with those of quarks.
The recent refined measurements of the proton and neutron
structure functions
revealed the violation of the Gottfried sum rule
and indicated
an excess of $d\bar d$ quark pairs
over $u \bar u$ quark pairs in the proton sea \cite{NMC91}.
The observation of the violation of 
the Gourdin-Ellis-Jaffe sum rule \cite{EJSR}
in polarized DIS experiments \cite{EMC,SMC,E142,E143}
gave rise to the proton ``spin crisis'' or ``spin puzzle'' and
triggered a vast number of theoretical and experimental investigations
on the spin content of the nucleons. There has also been significant
progress in the theory of the QCD corrections to the various
parton sum rules; e.g.,  the
generalized Crewther relation connects the observables in $e^+ e^-$
annihilation and the Bjorken \cite{BSR} and GLS sum rules in DIS,
providing a precision test of the standard
model with no scale or scheme ambiguities \cite{Bro95}.

All of the above mentioned parton sum rules are related
to the quark momentum distributions $q(x)$ and helicity
distributions $\Delta q(x)$, two of the fundamental
distributions which characterize the state of quarks in the
nucleon at leading twist. The above
two quark distributions are related to the
vector quark current $\bar q \gamma^{\mu} q$ and
the axial quark current
$\bar q \gamma^{\mu} \gamma^5 q$ respectively. There is
another fundamental distribution,
the quark transversity distribution $\delta q(x)$
which is related to the matrix elements of the tensor quark
current $\bar q \sigma^{\mu\nu} i \gamma^5 q$ \cite{h1}.
Unfortunately, there is still no suggestion of a basic
parton sum rule in analogy to the Bjorken sum rule
for the quark transversity distributions.
However, it has been recently shown that there is a
relation \cite{Ma97} which connects the quark transversity distributions
to the quark helicity distributions $\Delta q(x)$ and the quark model
spin distributions:
\begin{equation}
\Delta q_{QM}(x)+\Delta q(x) = 2 \delta q(x),
\label{mss}
\end{equation}
where $\Delta q_{QM}(x)$ is the quark spin distributions
as defined in the quark model or in the nucleon rest frame
\cite{Ma91b,Bro94,Ma96,Sch97}.
One can use this relation to measure
the quark model spin distributions once the quark helicity distributions
and the quark transversity distributions are both measured.
In this paper we will show that one can
connect the quark tensor charges
to the measured quantities $g_A/g_V$, $\Gamma_p$,
$\Gamma_n$ and several quantities with limited uncertainties
by combining the relation eq.~(\ref{mss}) with
the parton sum rules for the quark helicity
distributions.

The spin-dependent structure functions for the proton and
the neutron, when expressed in terms of the quark helicity
distributions $\Delta q(x)$, should read
\begin{equation}
g_1^p(x)=\frac{1}{2}[\frac{4}{9}(\Delta u(x) +\Delta \bar u(x))
+\frac{1}{9}(\Delta d(x) +\Delta \bar d(x))
+\frac{1}{9}(\Delta s(x) +\Delta \bar s(x))],
\label{ejsp}
\end{equation}
\begin{equation}
g_1^n(x)=\frac{1}{2}[\frac{1}{9}(\Delta u(x) +\Delta \bar u(x))
+\frac{4}{9}(\Delta d(x) +\Delta \bar d(x))
+\frac{1}{9}(\Delta s(x) +\Delta \bar s(x))].
\label{ejsn}
\end{equation}
The measured Gourdin-Ellis-Jaffe integrals
$\Gamma^p=\int_0^1 {\rm d} x g_1^p(x)$
and $\Gamma^n=\int_0^1 {\rm d} x g_1^n(x)$ from
polarized DIS experiments \cite{EMC,SMC,E142,E143}
have been
found to be in conflict with the corresponding sum rules
\cite{EJSR} based on assumptions of zero strangeness,  zero gluon spin
contribution, and SU(3) symmetry for the octet baryons.
The quark axial charge or the quark helicity
related to the axial quark current
$\bar q \gamma^{\mu} \gamma^5 q$ is
expressed by $\Delta Q=\int_0^1{\rm d} x [\Delta q(x)+\Delta \bar
q(x)]$.
Combining eqs.(\ref{ejsp}) and (\ref{ejsn}) one
obtains the sum of the quark axial charges (or quark helicities)
for the light flavors
\begin{equation}
\Delta U +\Delta D=
\frac{18}{5}(\Gamma^p+\Gamma^n)
-\frac{2}{5}\Delta S,
\label{ejspn}
\end{equation}
from which we know that the sum of quark helicities
$\sum \Delta Q$
including the strangeness contribution should be
\begin{equation}
\sum \Delta Q=\Delta U +\Delta D+\Delta S=
\frac{18}{5}(\Gamma^p+\Gamma^n)
+\frac{3}{5}\Delta S,
\label{sum}
\end{equation}
where the quantities $\Gamma^p$, $\Gamma^n$,
and $\Delta S$ can be measured independently
in different experiments.
The Bjorken sum rule defined by
\begin{equation}
\Gamma^p-\Gamma^n=
\frac{1}{6}
(\Delta U-\Delta D)=\frac{1}{6}\frac{g_A}{g_V},
\label{bjs}
\end{equation}
where $g_A/g_V$ is determined from the neutron $\beta$
decay, is a more basic result
and has been found to be valid with the observed
values of $\Gamma^p$ and $\Gamma^n$ within experimental
uncertainties by taking into account
QCD radiative corrections \cite{SMC,E142,E143}.

The tensor charge, defined as
$\delta Q=\int_0^1 {\rm d} x [ \delta q(x) -\delta \bar q(x)]$,
is chiral-odd due to the charge conjugation
properties of the tensor current $\bar q \sigma^{\mu\nu} i \gamma^5 q$.
Therefore the quark tensor charge $\delta Q$ and the
quark axial charge $\Delta Q$ have different chiral parities.
The quark helicity distributions, $\Delta q(x)$ and
$\Delta \bar q(x)$, and the quark
transversity distributions, $\delta q(x)$ and $\delta \bar q(x)$,
should be measured for quarks and
anti-quarks separately in applying Eq.~(\ref{mss}).
In order to get the tensor charge for each flavor, we must first
isolate
eq.~(\ref{mss})
for both quarks and anti-quarks of each flavor, and integrate.
In practice, one expects the antiquark contributions to be small. For
example, the anti-quark contributions 
to $\Delta Q$ and $\delta Q$ are zero
in the meson-baryon fluctuation model \cite{Bro96} and in a
broken-U(3) version of the chiral quark model \cite{Che95}.
There has been an explicit measurement of the helicity distributions
for the individual $u$ and $d$ valence and sea quarks by the Spin
Muon Collaboration (SMC) \cite{NSMCN}.  The helicity distributions
for the $u$ and $d$ anti-quarks are consistent with zero in agreement
with the results of the light-cone meson-baryon fluctuation model of
intrinsic $q \bar q$ pairs.

We thus can assume
that the anti-quark contributions are negligible.
We thus obtain, combining eqs.~(\ref{mss})
and (\ref{bjs}),
\begin{equation}
\delta U -\delta D=\frac{1}{2}[
(\Delta U-\Delta D)+(\Delta U_{QM}-\Delta D_{QM})],
\label{sum10}
\end{equation}
where the first term in the right side satisfies the Bjorken
sum rule and
the second term satisfies a Bjorken-like sum rule in which
one can approximate the quantity $\Delta U_{QM}-\Delta D_{QM}$
by the non-relativistic value 5/3 for the
naive quark model.
Therefore
we have a Bjorken-like sum rule for the isovector tensor charge
\begin{equation}
\delta U -\delta D=
\frac{1}{2}(\frac{g_A}{g_V}+\frac{5}{3} c_1)
\label{sum1}
\end{equation}
where $g_A/g_V$ might be the value from the neutron $\beta$
decay or $g_A/g_V=6 (\Gamma^p-\Gamma^n)$ from eq.~(\ref{bjs})
and $c_1$ is an unknown correction factor reflecting the deviation
from the naive quark model value
$\Delta U_{QM}-\Delta D_{QM}=5/3$
and might range from 0.9 to 1.
Similarly, combing eqs.~(\ref{mss}) and (\ref{sum}), we
obtain the second sum rule for the isoscalar tensor charge
\begin{equation}
\delta U +\delta D=
\frac{1}{2}[
(\Delta U+\Delta D)+(\Delta U_{QM}+\Delta D_{QM})]
=\frac{9}{5}(\Gamma^p+\Gamma^n)
-\frac{1}{5}\Delta S+\frac{1}{2} c_2,
\label{sum2}
\end{equation}
where $c_2$ is another unknown correction factor
reflecting the deviation
from the naive quark model value
$\Delta U_{QM}+\Delta D_{QM}=1$
and might range from 0.75 to 1.

 From eqs.~(\ref{sum1}) and (\ref{sum2}),
we can predict the quark tensor charges $\delta U$
and $\delta D$ by use of the measurable quantities $\Gamma_p$, $\Gamma_n$,
$g_A/g_V$ and $\Delta S$, and the correction factors $c_1$ and
$c_2$ with limited uncertainties.
The quantities $\Gamma^p$ and $\Gamma^n$ at several
different $Q^2$ have been
measured from polarized DIS experiments \cite{EMC,SMC,E142,E143}, and 
$\Delta S$ has also been
extracted from analysis of the polarized DIS data
and it might range from about -0.01 \cite{Bro96}
to -0.13 \cite{Ell95b}.
The value of $\Delta S$ from those analysis is sensitive
to the assumption of SU(3) symmetry. It would be
better to measure $\Delta S$
from other independent processes and there have been suggestions
for this purpose \cite{Lu95,Ell95}.
Nevertheless, we notice that the predicted
values of $\delta U$ and $\delta D$ are not sensitive to
$\Delta S$.
In case $g_A/g_V=6 (\Gamma^p-\Gamma^n)$ is adopted (we denote
case 1), for $\Gamma^p({\rm E143})=0.127$ and
$\Gamma^n({\rm E143})=-0.037$ at $\langle Q^2 \rangle=3$ GeV$^2$
\cite{E143},
we have
\begin{equation}
\begin{array}{clcr}
\delta U=0.89 \to 1.01;\\
\delta D=-0.28 \to -0.39,
\label{con1}
\end{array}
\end{equation}
and for $\Gamma^p({\rm SMC})=0.136$ and
$\Gamma^n({\rm SMC})=-0.063$ at $\langle Q^2 \rangle=10$ GeV$^2$
\cite{SMC},
we have
\begin{equation}
\begin{array}{clcr}
\delta U=0.93 \to 1.04;\\
\delta D=-0.34 \to -0.46.
\label{con2}
\end{array}
\end{equation}
Combining the above two constraints and
taking into account further the uncertainties (0.05)
introduced by the data,
we have
\begin{equation}
\begin{array}{clcr}
\delta U=0.84 \to 1.09;\\
\delta D=-0.23 \to -0.51.
\end{array}
\end{equation}
In case the value $g_A/g_V=1.2573$ from neutron $\beta$ decay is
adopted (we denote case 2),
we obtain
\begin{equation}
\begin{array}{clcr}
\delta U=0.94 \to 1.06;\\
\delta D=-0.36 \to -0.48
\label{con1b}
\end{array}
\end{equation}
corresponding to eq.~(\ref{con1}) and
\begin{equation}
\begin{array}{clcr}
\delta U=0.96 \to 1.07;\\
\delta D=-0.34 \to -0.46
\label{con2b}
\end{array}
\end{equation}
corresponding to eq.~(\ref{con2}).
We notice that the difference between eqs.~(\ref{con1b}) and
(\ref{con2b}) is much smaller than that between eqs.~(\ref{con1}) and
(\ref{con2}). This indicates the sensitivity to the quantity
$g_A/g_V$ used in the sum rule (\ref{sum1}).
Combining the constraints (\ref{con1b}) and (\ref{con2b})
and taking into account also the uncertainties 0.05,
we obtain
\begin{equation}
\begin{array}{clcr}
\delta U=0.89 \to 1.11;\\
\delta D=-0.29 \to -0.53.
\end{array}
\end{equation}
Further progress in the precision of the data and
in the knowledge of the correction factors can further
constrain the results.
Therefore the predicted $\delta U$ and $\delta D$ are within
limited ranges from the two sum rules eqs.~(\ref{sum1}) and
(\ref{sum2}).

\begin{table}[ht]
\begin{center}
TABLE 1 \\
\vspace{.6truecm}
\begin{footnotesize}
\begin{tabular}{|c|c|c|c|c|}
\hline\ru1 {\rm Name of work}
& {$\delta U$} & {$\delta D$}
& {$\delta U- \delta D$}
& {$\delta U+ \delta D$}\\
\hline\ru1
{\rm Case 1 of this work}
& {$0.84 \to 1.09$}
& {$-0.23 \to -0.51$}
& {$1.24 \to 1.43$}& {$0.51\to 0.69$} \\
\hline\ru1
{\rm Case 2 of this work}
& {$0.89 \to 1.11$}
& {$-0.29 \to -0.53$}
& {$1.38 \to 1.46$}& {$0.51 \to 0.69$} \\
\hline\ru1
{\rm Light-cone quark model \cite{Sch97}}
& {1.167}
& {-0.292}
& {1.458}& {0.875} \\
\hline\ru1
{\rm QCD sum rule \cite{He95}}
& {$1.33 \pm 0.53$}
& {$0.04\pm 0.02$}
& {$1.29\pm 0.51$}& {$1.37\pm 0.55$} \\
\hline\ru1
{\rm Chiral soliton model \cite{Kim96}}
& {1.12}
& {-0.42}
& {1.54}& {0.70} \\
\hline\ru1
{\rm Chiral chromodielectric model \cite{Bar97}}
& {0.969}
& {-0.250}
& {1.219}& {0.719} \\
\hline\ru1
{\rm Spectator model \cite{Mul97}}
& {1.218}
& {-0.255}
& {1.473}& {0.963} \\
\hline\ru1
{\rm Lattice QCD \cite{lQCD}}
& {$0.84$}
& {$-0.23$}
& {1.07}& {0.61} \\
\hline\ru1
{\rm Non-relativistic limit}
& {$\frac{4}{3}=1.333$}
& {$-\frac{1}{3}=-0.333$}
& {$\frac{5}{3}=1.667$}& {1} \\
\hline\ru1
{\rm Ultra-relativistic limit \cite{Bro94,Sch97}}
& {$\frac{2}{3}=0.667$}
& {$-\frac{1}{6}=-0.167$}
& {$\frac{5}{6}=0.833$} & {$\frac{1}{2}=0.5$} \\
\hline
\end{tabular}
\end{footnotesize}
\end{center}
\end{table}

We list 
in Table~1 our predictions of the
quark tensor charges $\delta U$ and $\delta D$
and the values of the two sums (\ref{sum1}) and (\ref{sum2}).
There have been a number of calculations of the
quark tensor charges $\delta U$ and $\delta D$, and
a comparison of our
results with several existing predictions
\cite{Sch97,He95,Kim96,Bar97,Mul97,lQCD}
is also made 
in Table~1.
 From the table we notice the significant
difference between the predictions.
One interesting feature we notice is that
the value of the first sum (i.e., the isovector tensor charge
$\delta U-\delta D$)
in our work
is consistent with most other predictions except
the lattice QCD result, whereas the value of the
second sum (i.e., the isoscalar tensor charge
$\delta U+\delta D$) is small and
only consistent
with the lattice QCD result \cite{lQCD}.
The small $\delta U +\delta D$ in our work
seems to be more
reasonable in analogy to the unexpected small
quark helicity sum $\Delta U+\Delta D$
which gave rise to the ``spin puzzle''.
It is also supported by a Skyrme model analysis in which
$\delta U+\delta D$ is of the order of $1/N_c$ relative to
$\delta U-\delta D$ in the large-$N_c$,
SU(3)-symmetric limit \cite{Oln93}.
 From another point of view, a small $\delta U+\delta D$
can be naturally understood within a framework of the
SU(6) quark spectator model \cite{Ma96} {\it plus} the baryon-meson
fluctuation model \cite{Bro96}: the flavor asymmetry
between the Melosh-Wigner rotation factors for the
$u$ and $d$ quarks will cause a reduction of
$\delta U+\delta D$ relative to the flavor symmetric case \cite{Sch97},
and a further reduction comes from an additional negative
contribution to $\delta D$ due to the intrinsic $d \bar d$
fluctuations related to the Gottfried sum rule violation.
The future experimental measurements of $\delta U$ and
$\delta D$ can test the above predictions and reveal
more information of the quark-gluon structure of the nucleon if
the measured values will be out of the predicted ranges.

We should mention that since there is
no fundamental physical tensor
current, 
the proposed sum rules have then the correction coefficients,
i.e., they
are not exact.
We have neglected the contributions from anti-quarks, gluons,
$Q^2$ dependence due to higher twist effects,
and different evolution behaviors between $\Delta Q$ and
$\delta Q$ in the above analysis.
In principle
the corrections
due to these sources can be further
taken into account
from theoretical and experimental progress and
they should be topics for later study.
We indicate that the contributions due to
gluons or sea quarks might be canceled
in $\delta U-\delta D$ and $\Delta U_{QM}-\Delta D_{QM}$,
in analogy to the situation of $\Delta U-\Delta D$ \cite{Sch97}.
Therefore the first sum rule (\ref{sum1}) might be more basic
than the second one (\ref{sum2}), and that is also
why we adopted a small uncertainty ($0.9 \to 1$)
for the correction
factor $c_1$ compared to $c_2$ with a large uncertainty
($0.75\to 1$) due to the possible negative contribution
from the sea quarks \cite{Bro96}.

One of the known constraints for the quark transversity
distributions is Soffer's inequality \cite{Sof95}:
\begin{equation}
q(x) + \Delta q(x) \ge 2\big |\delta q(x)|,
\label{Sie}
\end{equation}
which is valid for each flavor,
likewise for antiquarks. We need to check whether our predicted
values for $\Delta U$ and $\delta D$ satisfy this
inequality, if we neglect antiquark contributions as was explained
before.
At a first sight one may have doubt since
$\delta D$ can be -0.5 from Table 1, whereas the measured
$\Delta D$ is around -0.35 and the integrated
$\int_0^1 [{\rm d} x] d(x)$ for valence quark is only 1. However,
one should take into account the $d$ sea quarks
for the first term of (\ref{Sie}). From
the Gottfried sum rule violation \cite{NMC91} we know that the excess
of $d \bar d$ over $u \bar u$ should be of the order 0.15
and in principle there could be also unlimited numbers
of extrinsic sea quarks in the nucleon sea \cite{Bro96}. 
Thus there is
no difficulty to satisfy the Soffer's inequality for
the values of $\delta U$ and $\delta D$ 
predicted from the two sum rules (\ref{sum1})
and (\ref{sum2}).

\vspace{0.5cm}
\begin{figure}[htbp]
\begin{center}
\leavevmode {\epsfysize=13.0cm \epsffile{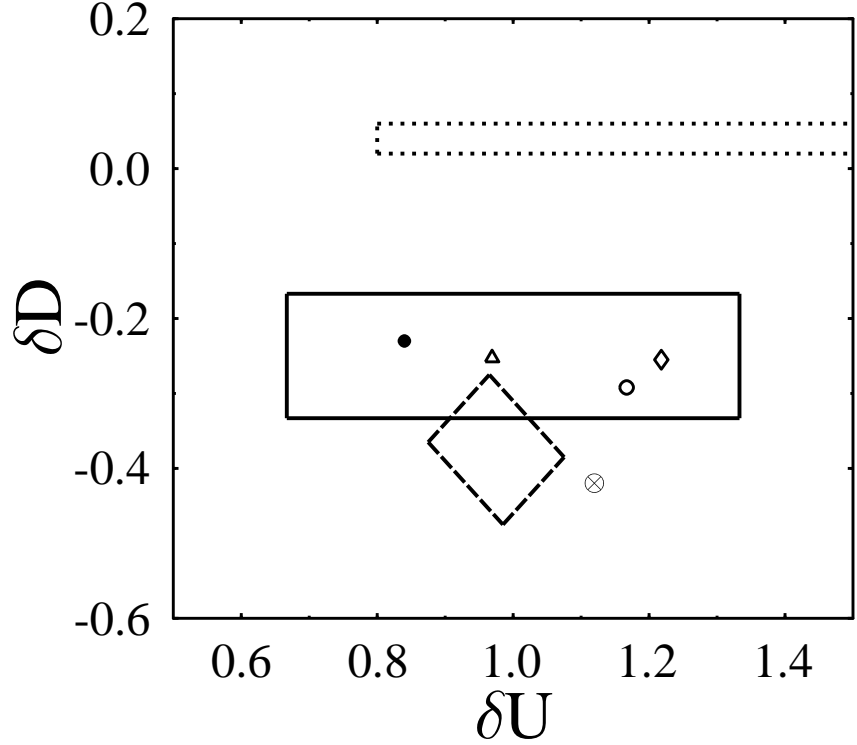}} 
\end{center}
\caption[*]{\baselineskip 13pt 
The predictions of the quark tensor charges
$\delta U$ and $\delta D$.
The markers are predictions from several models:
the light-come quark model $\circ$ \cite{Sch97}, 
the chiral soliton model $\otimes$ \cite{Kim96},
the chiral chromodielectric model $\triangle$ \cite{Bar97}, 
the spectator model $\diamond$
\cite{Mul97}, and lattice QCD $\bullet$ \cite{lQCD}.
The solid box represents the range within
the non-relativistic and ultra-relativistic limits
of the simple three quark light-cone model \cite{Bro94,Sch97}, 
the dotted box represents the prediction
from the QCD sum rule \cite{He95}, and the dashed box
represents the range predicted from the two sum rules
Eqs.~(\ref{sum1}) and (\ref{sum2}).}  
\label{bmsf1}
\end{figure}

We also list in Table 1 the values of $\delta U$ and $\delta D$
in the non-relativistic and ultra-relativistic limits 
\cite{Bro94,Sch97} of the simple three quark light-cone model. 
The predicted quark tensor charges 
$\delta U$ and $\delta D$ listed in Table 1 
are also presented in Fig.~1.
It is
interesting to note that the tensor charges still have
finite values in the ultra-relativistic limit, compared
to the corresponding case of vanishing axial charges \cite{Bro94}.
We also notice that the predicted values for
$\delta U$, $\delta U -\delta D$, and $\delta U+\delta D$
are within the values between the two limits, whereas
the predicted $\delta D$ may have an additional negative
contribution beyond the naive quark model. This is 
similar to the case of the axial charges discussed
in Ref.~\cite{Bro96}. 
Unlike most other predictions, the QCD sum rule
analysis \cite{He95,Jin97} predicted a shift of $\delta D$ beyond
the quark model limits in an opposite direction.
Thus any evidence of the
measured $\delta D$ beyond the range $-1/6 \to -1/3$ will
be useful to confirm contribution from the intrinsic $d$
sea quarks
predicted in Refs.~\cite{Ma97,Bro96} or other new physics.

In summary, we proposed in this paper two new sum rules,
based on a known relation connecting
the quark transversity distributions to the
quark helicity distributions and the quark model spin distributions,
and on the sum rules for the quark helicity distributions.
Though the two sum rules are simple, they are
useful to predict the values of
the quark tensor charges
$\delta U$ and $\delta D$
from the measured quantities of $\Gamma^p$, $\Gamma^n$,
$g_A/g_V$ and
$\Delta S$, and two model correction factors with limited
uncertainties.
We also predicted a small value for the sum of the
quark tensor charges compared to most other predictions,
and this seems to be reasonable in analogy
to the unexpected small quark helicity sum
which gave rise to the proton ``spin puzzle''.

{\bf Acknowledgments: }
We are greatly indebted to Stan J. Brodsky
for his inspiring and valuable discussions and suggestions
during this work.
We are also very grateful to Jacques Soffer for his
comments on the paper.

\end{document}